# Probing hidden Mott gap and incommensurate charge modulation on the polar surfaces of PdCrO$_2$

Chenhaoping Wen[1,*], Tian Qin[1,*], Xuefeng Zhang[1,*], Mingxin Zhang[1,*], Shiwei Shen[1], Zhongzheng Yang[1], Yanpeng Qi[1,2,3,†], Gang Li[1,2,‡], Shichao Yan[1,2,§]

[1] *School of Physical Science and Technology, ShanghaiTech University, Shanghai 201210, China*

[2] *ShanghaiTech Laboratory for Topological Physics, ShanghaiTech University, Shanghai 201210, China*

[3] *Shanghai Key Laboratory of High-resolution Electron Microscopy, ShanghaiTech University, Shanghai 201210, China*

[*] *These authors contributed equally to this work*
[†] *Email*: *qiyp@shanghaitech.edu.cn*
[‡] *Email*: *ligang@shanghaitech.edu.cn*
[§] *Email*: *yanshch@shanghaitech.edu.cn*

**Abstract**

Here we report a combined study of low-temperature scanning tunneling microscopy (STM) and dynamical mean-field theory (DMFT) on PdCrO$_2$, a delafossite metal with an antiferromagnetic order below ~37.5 K. First, on the CrO$_2$-terminated polar surface we detect a gap-like feature both below and above the Néel temperature. The DMFT calculations indicate that this gap is opened due to the strong correlations of Cr-3d electrons, suggesting the hidden Mott nature of the gap. Then, we observe two kinds of Pd-terminated polar surfaces. One is a well-ordered Pd surface with the Fermi-surface-nesting-induced incommensurate charge modulation, while the other one is a reconstructed Pd surface with the individual nano-scale non-periodic domain structures. On the well-ordered Pd surface, the interference between the incommensurate charge modulation and the atomic lattice forms the periodic moiré pattern. Our results provide important microscopic information for fully understanding the correlated electronic properties of this class of materials.

Correlated electronic phases often emerge accompanied by the antiferromagnetic (AFM) order, such as the unconventional superconductivity in cuprates and iron-based superconductors [1-3]. Fully understanding these correlated electronic phases is challenging because of the complicated interplay among different electronic orders, such as antiferromagnetism, nematicity, and superconductivity [1-3]. Exploring the interaction between the conduction electrons and the localized magnetic moments in a simpler material would be helpful to gain a deeper understanding of the correlated electronic systems. In this respect, the delafossite compound $PdCrO_2$, a natural heterostructured material with individual layers containing itinerant electrons and localized spins, meets this condition [4-9].

Sharing the same crystal structure as the typical delafossite materials, $PdCrO_2$ consists of the alternative stacking of the triangular Pd layer and $CrO_2$ layer [8,10-12], in which the Cr and Pd atoms offer 3d and 4d electrons with distinct responses to charge and spin fluctuations. $PdCrO_2$ shows metallic conductivity due to the Pd $4d^9$ electrons, and the localized spins of $Cr^{3+}$(S = 3/2) lead to the onset of AFM order with the Néel temperature ($T_N$) of ~37.5 K [13]. According to the neutron scattering measurements, the spin order in $PdCrO_2$ has a 120˚ AFM spin structure with √3 by √3 periodicity [14-16], but the unconventional anomalous Hall effect found in $PdCrO_2$ suggests that it may have a chiral spin structure at low temperature [17].

For the electronic structure of $PdCrO_2$, both the quantum oscillations and angle-resolved photoemission spectroscopy (ARPES) measurements show that the Fermi surface of $PdCrO_2$ is a single hexagon above $T_N$, which mainly originates from the Pd electrons [5,18-20]. This hexagonal Fermi surface is backfolded into the antiferromagnetic Brillouin zone below $T_N$ [18-20]. More recently, the combined density functional theory with dynamical mean-field theory (DFT+DMFT) calculations suggest that the Cr layer shows a Mott gap behavior that is driven by the Cr 3d local moments [4,21,22]. So far, the microscopic electronic information on the Pd and the $CrO_2$ layers, which is great essential for fully understanding the electronic behaviors of this material, is still lacking.

Here we use low-temperature scanning tunneling microscopy (STM) and DFT+DMFT to systematically study $PdCrO_2$. Our STM data provide complete information for both the $CrO_2$- and Pd-terminated polar surfaces. On the $CrO_2$-terminated surface, we observe a gap-like feature both below and above $T_N$. Our DMFT reveals that this gap is opened for the Cr 3d electrons driven by electronic correlations, leaving the pure Fermi surface originating from the Pd layer. On the Pd-terminated surface, we observe a Fermi surface nesting induced two-dimensional incommensurate charge modulation, which is due to the spontaneous symmetry-breaking of the correlated Fermi surface nicely consistent with our DMFT-based random phase approximation (RPA) calculations. In addition, we also find the reconstructed Pd surface with nano-scale domains consisting of the locally-ordered Pd atoms. Our results provide detailed microscopic information for the polar surfaces of $PdCrO_2$.

Single crystals of PdCrO$_2$ were grown with a similar procedure as reported in Ref. [10,23]. STM experiments were performed with a Unisoku 1200 low-temperature STM. PdCrO$_2$ single crystals were cleaved at 77 K with the pressure under $2 \times 10^{-10}$ Torr, and then were immediately transferred into the STM head for measurements at 4.3 K. Chemically etched tungsten tips were used for the STM measurements. STS measurements were done by using a standard lock-in technique with a 3 mV modulation at the frequency of 914 Hz. For the theoretical understanding, we constructed a model Hamiltonian consisting of Pd 4d and Cr 3d orbitals by using the Wannier90 package [24] and performed DMFT calculations [25,26] with our home-made code *PACS*. We treated the Cr 3d orbitals as correlated and set the on-site Hubbard interaction $U = 3$ eV and Hund coupling $J = 0.7$ eV. Our DMFT calculations were performed in the paramagnetic phase with a temperature $T \sim 290$ K. Continuous-time quantum Monte Carlo [27] was used as the impurity solver.

Due to the quasi-two-dimensional crystal structure of PdCrO$_2$, there are Pd and CrO$_2$ surface terminations after cleaving [Figs. 1(a) and 1(b)]. Figures 1(c) and (e) are the large-area STM topographies taken on the CrO$_2$- and the Pd-terminated polar surfaces, respectively. As shown in the line profile across the atomic steps [insets of Figs. 1(c) and 1(e)], both the two surfaces show clean and sharp atomic steps with a similar step height (~6 Å), indicating that both steps contain one complete CrO$_2$-Pd or Pd-CrO$_2$ unit along the $c$ direction. On the CrO$_2$ surface, we observe the triangular atomic lattice with a lattice constant of ~2.9 Å and a few atomic defects [Fig. 1(d)]. On the Pd surface, except for the atomic structure, we also find the hexagonal patterns with a larger real-space period than the atomic lattice [as indicated by the yellow dashed hexagons in Fig. 1(f)], making the Pd surface distinct from the CrO$_2$ surface.

We next investigate the two surface terminations separately. First, we perform scanning tunneling spectroscopy (STS) measurements on the CrO$_2$ surface to reveal its electronic structure. At 4.3 K, the d$I$/d$V$ spectra taken on a defect-free region of the CrO$_2$ surface show a gap-like feature at the Fermi level ($E_F$). This gap-like feature is homogenous on the CrO$_2$ surface, and it can barely be affected by the atomic impurities on the surface (see Supplemental Material for the d$I$/d$V$ spectra taken near the atomic impurities, Fig. S1 [28]). Below and beyond the gap, we observe two electronic peaks locating at −0.43 V and 0.7 V. Interestingly, we are still able to detect the finite density of states (DOS) within the energy gap [inset of Fig. 2(b)], which could be contributed by the electrons from the underneath Pd layer.

To understand the nature of the gap-like feature on the CrO$_2$ surface, we calculate the correlated electronic structure of PdCrO$_2$ by using DMFT. Figure 2(c) shows the spectral function $A(\mathbf{k}, \omega)$ of PdCrO$_2$ in the $k_z = 0$ plane. Along the Γ-K and Γ-M direction, a single band with large Fermi velocity crosses $E_F$, which mainly consists of the Pd 4d electrons. Compared to the DFT [22], the electronic states of the Cr 3d electrons, however, have been pushed away from $E_F$ by the on-site Hubbard interaction. The lower (LHB) and the upper Hubbard band (UHB) stay below −0.4 eV and above 0.1 eV, respectively. The Mott gap is of size ~0.5 eV, which is highly consistent with our STS measurements. Our paramagnetic DMFT results indicate that it is the electronic correlations and local fluctuations of the Cr 3d electrons rather than the long-range orders that open the gap, which is compelling evidence for the Mott physics that equally applies to the temperatures below and above $T_N$.

To further confirm this conclusion, we investigate the evolution of the gap size as a function of temperature from below to above the $T_N$. If the insulating gap is induced by the long-range AFM order, it should collapse when the temperature is above $T_N$. Figure 2(d) shows the temperature dependence of the d$I$/d$V$ spectrum taken on the $CrO_2$ surface. The insulating gap can still be seen even when the temperature is increased up to ~45 K. The Néel temperature of $PdCrO_2$ is ~37.5 K, and the electronic gap is preserved across the AFM phase transition. This result nicely agrees with our DMFT calculations that the Mott gap is opened even when the system is in the paramagnetic state. We also note that the Mott gap on the $CrO_2$ surface can barely be tuned by the surface K doping, as shown in Figs. 2(e) and 2(f). Upon doping the K atoms onto the surface, we only observe a slight downshift of the peaks of the occupied states, and we find no in-gap states appeared in the Mott gap. This might be because the $CrO_2$ surface is a negatively charged polar surface, and the K atoms cannot dope electrons into it.

Having characterized the $CrO_2$ surface, we turn to the Pd-terminated surface. Different from the atomic lattice on the $CrO_2$ surface, the Pd surface has a regular lattice with ~4.0 Å periodicity, which is ~1.38 times the lattice constant on the $CrO_2$ surface (see Supplemental Material for the comparison between the two surfaces, Fig. S2 [28]). The STM topographies taken on the Pd surface of $PdCrO_2$ change dramatically as changing the STM bias voltages (see Supplemental Material for the bias-voltage dependent STM topographies, Fig. S3 [28]). Apart from the regular lattice, the stripe-like patterns appear near the impurities in the STM topography taken with + 200 mV bias voltage [Fig. 3(a)]. The stripe-like patterns can usually spread to a distance up to ~10 nm. In the STM topography taken with the negative bias voltages [Fig. 3(b)], a superstructure with larger periodicity (~ 9.7 Å) appears.

To quantify the periodicities of the regular lattice and the superstructure on the Pd surface, we perform Fourier transform (FT) to the STM topographies taken on the Pd surface (see Supplemental Material for more FT images, Fig. S4 [28]). Figures 3(c) and 3(d) show the FT images of the topographies in Figs. 3(a) and 3(b), respectively. We find that there are three sets of wavevectors marked by the orange ($Q_0$), blue ($Q_1$) and green ($Q_{Bragg}$) circles. The $Q_{Bragg}$ corresponds to the atomic lattice on the Pd surface which has a period of ~2.9 Å. The $Q_0$ wavevector is along the $Q_{Bragg}$ direction while the $Q_1$ wavevector has an angle of 30° with the $Q_{Bragg}$ wavevector. The lengths of the $Q_1$ and $Q_0$ wavevectors are independent of bias voltages as shown in the line profiles in the FT images along the $Q_1$ and $Q_{Bragg}$ directions (see Supplemental Material, Fig. S5 [28]). The ratio between the $Q_1$ and $Q_{Bragg}$ wavevectors is ~0.74 ± 0.02, and the ratio between the $Q_0$ and $Q_{Bragg}$ wavevectors is ~0.30 ± 0.02. This indicates that the $Q_1$ wavevector corresponds to the regular lattice of 4.0 Å and the $Q_0$ wavevector corresponds to the large periodicity of ~ 9.7 Å. Such $Q_0$ and $Q_1$ wavevectors have also been recently reported on the Pd-terminated surface of $PdCoO_2$, where the $Q_1$ wavevector is related to the charge modulation pattern, and the $Q_0$ wavevector is due to the interference between the $Q_1$ and $Q_{Bragg}$ related lattices [29].

To confirm the origin of the charge modulation on the Pd surface, we calculate the correlated Fermi surface of $PdCrO_2$. According to the DMFT calculations, the Fermi surface is mainly contributed by the Pd layer [Fig. 3(g)], and it is almost a perfect hexagon, which is consistent with the previous ARPES measurements [5]. The parallel segments on the hexagonal Fermi surface can provide a strong nesting effect, as indicated by the red arrows of $q_1$, $q_2$, and $q_3$. We also calculate the charge susceptibility which is shown in Fig. 3(h) by using the RPA method based on the DMFT correlated electronic structure (see Supplemental Material for more details [28]). We find that the charge susceptibility maximum appears along the Γ-M direction, which has a ratio of ~0.74 compared to the reciprocal vector. This ratio matches the measured ratio between $Q_1$ and $Q_{Bragg}$, indicating that $Q_1$ is likely to be induced by the Fermi surface nesting, and the regular lattice on the Pd surface is the charge modulation pattern. The nesting-induced charge instability in $PdCrO_2$ starkly contrasts with the conventional ones, as it emerges from a two-step process that is peculiar to this system. The $CrO_2$ and Pd layers are geometrically separated and electronically loosely coupled, such that the joint Fermi surface is first renormalized by the electronic correlation of Cr and forms a nesting hexagon that triggers the charge instability in the second step. The emergent $Q_1$ wavevector in the FT images of the topographies offers a piece of new evidence for the novel correlation effect in this 3$d$-4$d$ hybrid compound. The d$I$/d$V$ spectra on the Pd surface exhibit a ~20 meV dip-like feature neat $E_F$, which could be the gap related to this charge modulation [Fig. 3(f)]. On the other hand, because the charge modulation on the Pd surface is incommensurate with the atomic lattice, it can create an interference pattern with the atomic lattice and form the moiré pattern with larger periodicity. The $Q_0$ wavevector corresponds to the moiré superlattice of the charge modulation and the atomic lattice. This interference effect can be verified by simply overlapping the charge modulation pattern and the atomic lattice to deduce the superposition pattern (see Supplemental Material, Fig. S6 [28]).

Interestingly, except for the well-ordered Pd surface, we also observe another Pd surface with reconstructed patterns. Figure 4(a) shows an atomic step between the reconstructed Pd surface and the $CrO_2$ surface, and the step height is ~ 4 Å. We can clearly see that there are nano-scale domains consisting of the semi-ordered Pd atoms [Fig. 4(b)]. These domains have different sizes and shapes, and are separated by dark boundaries [as marked by the yellow dashed lines in Fig. 4(c)]. The regular size of the domain ranges from 1 nm to 2 nm. The spatially-resolved d$I$/d$V$ spectra indicate that the DOS is almost unchanged on the domain boundaries [Fig. 4(d)]. This suggests that the reconstructed Pd surface remains ordered especially in a small scale, and the dark boundaries between the domains could be induced by the missing Pd atoms during the cleaving process. This may also indicate that Pd vacancies can suppress the incommensurate charge modulation on the Pd surface of $PdCrO_2$.

In summary, we report a systemic study of the polar surfaces of $PdCrO_2$. On the $CrO_2$-terminated polar surface, we detect a hidden Mott gap both below and above the Néel temperature. The DMFT calculations indicate that this hidden Mott gap is opened due to the electronic correlations and fluctuations of the Cr 3d electron local moments. On the well-ordered Pd surface, we detect an incommensurate charge modulation originating from the Fermi surface nesting, which is supported by our renormalized RPA calculations. The incommensurate charge modulation creates an interference pattern with the atomic lattice, forming the moiré superlattice. On the reconstructed Pd surface, we find the individual nano-size domains formed by the locally ordered Pd atoms. Our results provide important microscopic insights for understanding the correlated electronic phases in this class of delafossite metals.

**ACKNOWLEDGEMENTS**

S.Y. and G.L. acknowledge the financial support from the National Key R&D Program of China (Grant No. 2022YFA1402703) and the start-up funding from ShanghaiTech University. C.W. acknowledges the support from National Natural Science Foundation of China (Grant No. 12004250) and the Shanghai Sailing Program (Grant No. 20YF1430700). G.L. acknowledges Shanghai 2021-Fundamental Research Aera 21JC1404700 for financial support. X.Z. acknowledges the Postdoctoral Special Funds for Theoretical Physics of the National Natural Science Foundation of China (Grant No. 12147124). Part of the calculations was performed at the HPC Platform of ShanghaiTech University Library and Information Services, and at the School of Physical Science and Technology. Y.Q. acknowledges the National Key R&D Program of China (Grant No. 2018YFA0704300) and the National Natural Science Foundation of China (Grant No. 52272265, 11974246 and U1932217).

**Figure 1**

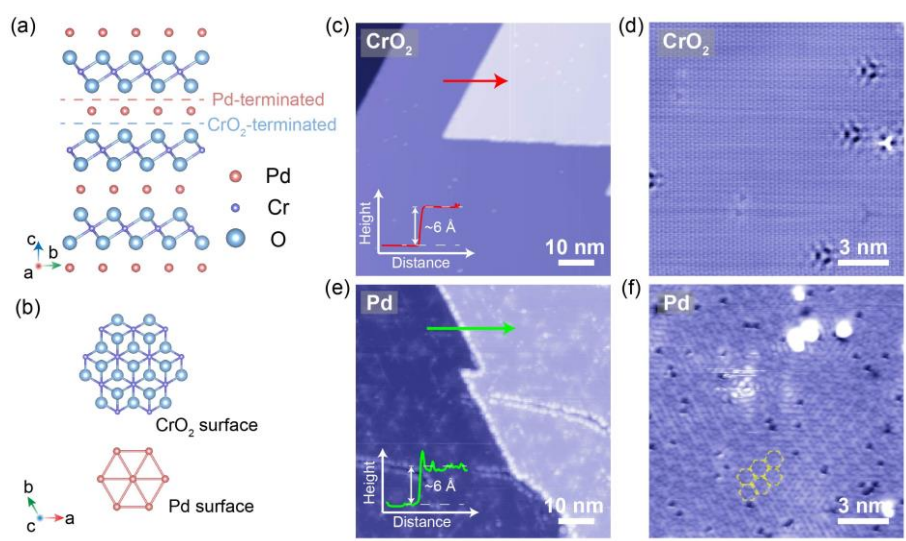

Fig. 1 (a) Crystal structure of PdCrO$_2$ from the side view. The red and blue dashed lines indicate the two typical cleavage planes. (b) Crystal structures of the CrO$_2$ (upper) and the Pd (bottom) surfaces from the top-down view. (c) Constant-current STM topography showing the atomic steps on the CrO$_2$ surface. The inset is the line profile along the red arrow. ($V_s$ = −500 mV, $I$ = 10 pA). (d) High-resolution STM topography on the CrO$_2$ surface. ($V_s$ = −500 mV, $I$ = 50 pA). (e) and (f) The same as that in (c) and (d) but measured on the Pd surface. The dashed yellow hexagons in (f) indicate the hexagonal superstructures. (e: $V_s$ = 200 mV, $I$ = 20 pA. f: $V_s$ = 50 mV, $I$ = 20 pA).

**Figure 2**

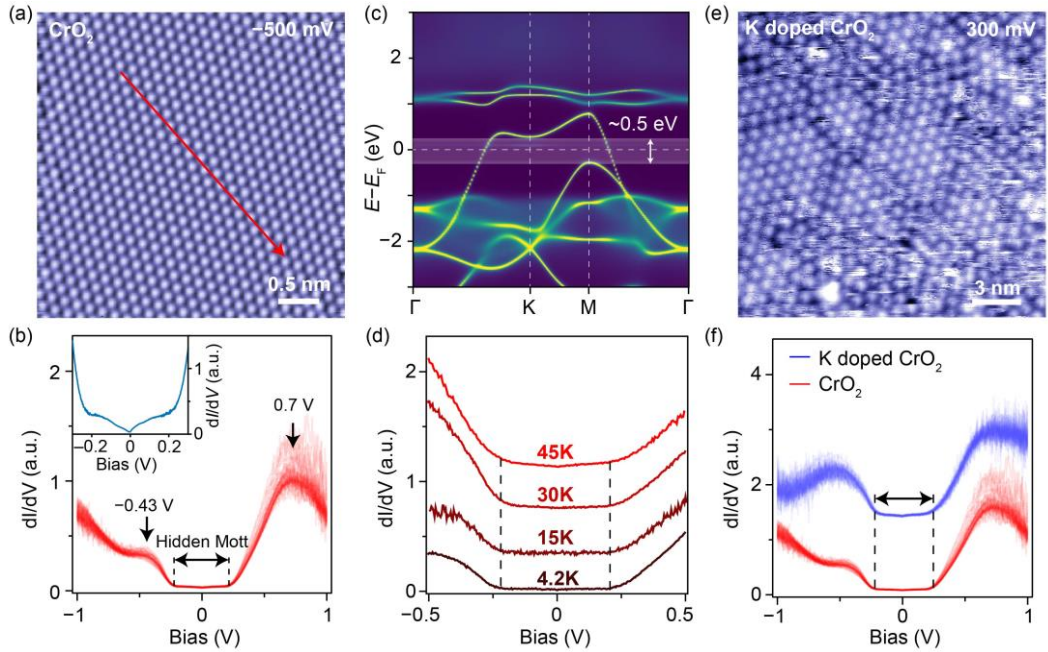

Fig. 2 (a) Atomic-resolution STM topography taken on the defect-free region of the $CrO_2$ surface. ($V_s = -500$ mV, $I = 20$ pA). (b) d$I$/d$V$ spectra taken along the red arrow in (a) (setup condition: $V_s = -1$ V, $I = 1$ nA). Inset: averaged smaller-range d$I$/d$V$ spectrum taken along the same line (setup condition: $V_s = -250$ mV, $I = 500$ pA). (c) Calculated spectral function $A(\mathbf{k}, \omega)$ along the high-symmetry lines in the reciprocal space. (d) Temperature-dependent d$I$/d$V$ spectra measured on the $CrO_2$ surface (setup condition: $V_s = 1$ V, $I = 1$ nA). The dashed lines indicate the position of the energy gap near $E_F$. (e) Atomic-resolution STM topography taken on the K-doped $CrO_2$ surface ($V_s = 300$ mV, $I = 5$ pA). (f) Comparison of the d$I$/d$V$ spectra taken on the pristine (red curve) and K-doped (blue curve) $CrO_2$ surface (setup condition: $V_s = 1$ V, $I = 1$ nA). The dashed lines and black arrow indicate the position of the energy gap at $E_F$.

**Figure 3**

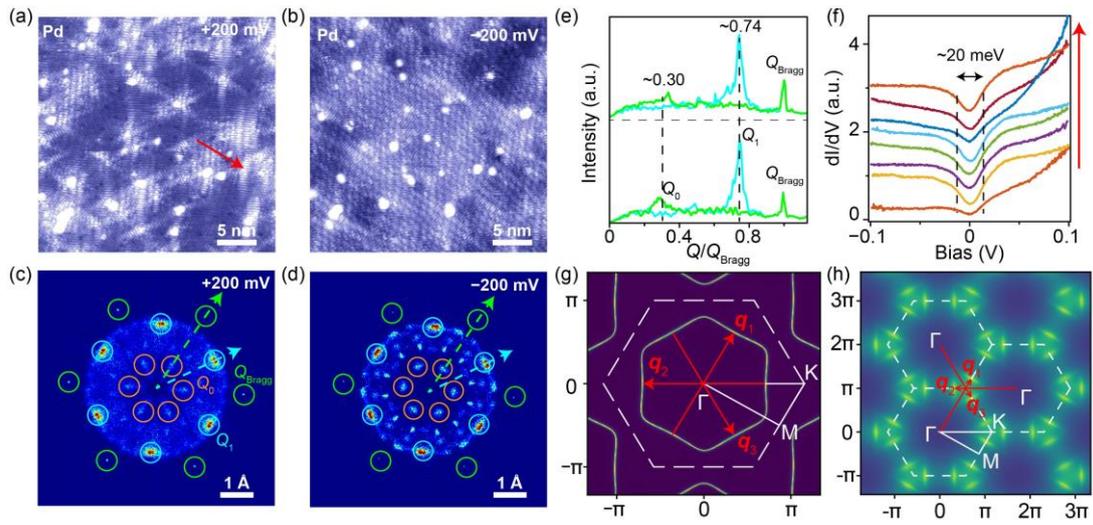

Fig. 3 (a) and (b) Constant-current STM topographies taken on the Pd surface with (a) 200 mV and (b) −200 mV bias voltages, respectively. (c) and (d) FT images of the STM topographies in (a) and (b). The green, blue, and orange circles correspond to the $Q_{Bragg}$, $Q_1$ and $Q_0$ wavevectors. (e) Line profiles of the FT images in (c) and (d) along the $Q_{Bragg}$ and $Q_1$ wavevectors. The dashed lines indicate the $Q_0$ and $Q_1$ peak positions. (f) The d$I$/d$V$ spectra taken along the red arrow in (a) (setup condition: $V_s$ = 100 mV, $I$ = 300 pA). (g) Calculated Fermi surface of PdCrO$_2$ in the Brillion zone along the $k_z$ = 0 plane. The red arrows indicate the nesting vectors. (h) Calculated charge susceptibility by using DMFT-based random phase approximation (RPA) method.

**Figure 4**

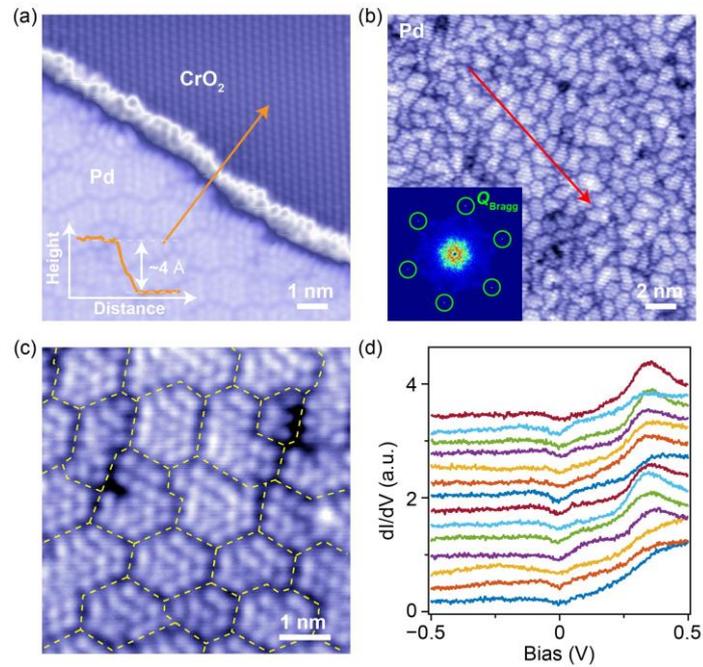

Fig. 4 (a) Constant-current STM topography measured near the atomic step between the reconstructed Pd surface and the $CrO_2$ surface. The inset is the line profile across the step along the orange arrow ($V_s = -500$ mV, $I = 20$ pA). (b) Constant-current STM topography taken on the reconstructed Pd surface ($V_s = -500$ mV, $I = 20$ pA). The inset shows its FT image. The green circles represent the $Q_{Bragg}$ wavevectors. (c) Zoom-in STM topography on the reconstructed Pd surface. The yellow dashed lines mark the positions of the non-periodic domain boundaries ($V_s = -500$ mV, $I = 20$ pA). (d) d$I$/d$V$ spectra measured along the red arrow in (b) (setup condition: $V_s = 500$ mV, $I = 150$ pA).